\documentstyle[11pt,a4,times,amssymb]{article}

\def\demi{{\textstyle {1\over2}}}
\def\Aphi{\chi}
\def\B{B}
\def\bB{\bar B}
\def\H{H}
\def\bH{\bar H}
\def\ephi{ \bar D^2 \bar \Phi + V'(\Phi) }
\def\bephi{ D^2 \Phi + V'(\bar \Phi) }
\def\dirac{ \not\! \partial}

\def\fV{\Psi_V}
\def\aV{\Psi^\flat_V}
\def\bV{B_V}

\def\fX{\Psi_X}
\def\aphi{\phi^\flat}

\def\da{{\dot\alpha}}

\let\nonu=\nonumber

\begin{document}
\bibliographystyle{perso}

\begin{titlepage}
\null \vskip -0.6cm
\hfill PAR--LPTHE 01--06

\hfill RUNHETC-2000-54

\hfill hep-th/0102075

\vskip 1.4truecm
\begin{center}
\obeylines

        {\Large	Supersymmetry with a Ghost Time}
\vskip 6mm
Laurent Baulieu$^\dagger$ and Marc Bellon
{\em Laboratoire de Physique Th\'eorique et Hautes Energies,
 Universit\'es Pierre et Marie Curie, Paris~VI
et Denis~Diderot,~Paris~VII}
 {\em and}
{\em
 $^{\dag}$  Dept. of Physics, Rutgers University, New Brunswick,
NJ~60637,~USA }

\end{center}

\vskip 13mm

\noindent{\bf Abstract}: The progress brought to the study of chiral
fermions and gauge theories by  quantization methods with a bulk time
suggests their usefulness in supersymmetric theories.  Using superspace
methods, we show how an explicitly supersymmetric version of such
quantization methods may be given.  \vfill

\begin{center}

\hrule \medskip
\obeylines
Postal address: %
Laboratoire de Physique Th\'eorique et des Hautes Energies,
 Unit\'e Mixte de Recherche CNRS 7589,
 Universit\'e Pierre et Marie Curie, bo\^\i te postale 126.
4, place Jussieu, F--75252 PARIS Cedex 05

\end{center}
\end{titlepage}

\section{Introduction}

Since the introduction of stochastic quantization for gauge theories
in~\cite{PaWu81}, its interest has been vindicated in a number of
publications (see~\cite{DaHu87} for a review). In recent
works~\cite{BaZw99,BaGrZw00,BaZw00}, the basic ideas of stochastic
quantization have been elaborated in a systematic approach, called bulk
quantization. It gives a central role to the introduction of a symmetry
of topological character.  The inclusion of fermionic fields becomes
possible.  Perturbatively, 
bulk quantization and the usual quantization methods are equivalent
because   the observables satisfy the same
Schwinger-Dyson equations in both approaches~\cite{BaZw00}. The interest
of bulk quantization is certainly at the non perturbative level. 

These recent developments make the application of bulk quantization to
supersymmetric field theories promising.  In this letter, we present the
simplest $N=1$ theory in four dimensions, the Wess--Zumino model of
interactions of chiral supermultiplets.

Using previous works~\cite{BaZw00}, we define an action on a space with an
additional non compact dimension $t$ that describes the bulk time.  The
correlation functions for equal $t$ define the correlations in the physical
theory.  The additional dimension does not take part in the Poincar\'e
group of symmetries.  Spinor fields
and the algebra of supersymmetry are thus the usual four-dimensional
ones.  It follows that one can use the superfield formalism
to efficiently describe the different supermultiplets and obtain an
invariant action. The supersymmetry is linearly realized and the algebra
closes off-shell, so that we do not have to worry on the preservation of
the supersymmetry by dynamical processes. As a side effect,  this
formalism  puts the auxiliary field of the
supermultiplet on the same level as the other components, with its own
dynamic.

The nice point in this construction is that the requirement of supersymmetry,
obtained through the superspace formalism, naturally 
introduces the  kernels acting on the equations of motion of the
four-dimensional theory, guaranteeing that each component has the proper
behavior. As it is the case in any field theory with fermions, the
difficulty of giving to  all equations  a  stochastic
interpretation  justifies to directly
postulate the topological field theory construction of the bulk theory.

We conclude this note by a short account of the extension to the case of gauge
invariant models: the gauge symmetry must be  enlarged in a superspace
formalism and we point to the difficulties in which would run too naive an
approach.  A full account of supersymmetric gauge theories will be given in a
forthcoming publication~\cite{BaBe01}.

\section{Wess--Zumino model}

In this model, the fields are grouped in supermultiplets which are
easily described by chiral superfields.  In view of our general principles,
these fields will depend on the space coordinates $x$, the anticommuting
superspace coordinates $\theta, \bar\theta$ and the ghost time $t$.  
\begin{equation}
 \Phi(t,x,\theta, \bar\theta),\qquad \bar D_{\dot\alpha} \Phi = 0.
\end{equation}
The model is then characterized by the superpotential $V(\Phi)$, which is
at most trilinear for renormalizable interactions.  The equation of motions
for $\Phi$ and its  antichiral conjugate $\bar\Phi$ are expressed in
superfields as:
\begin{eqnarray}\label{movWZ}
        0 &=& \bar D^2 \bar\Phi + V'(\Phi) = \bar D^2 \bar\Phi + m \Phi +
        \demi g \Phi^2, \nonu \\
        0 &=& D^2 \Phi + V'(\bar\Phi) .
\end{eqnarray}
Formally, these equations are very similar to the Dirac equations, with
$D^2$ and $\bar D^2$ playing the role of the chiral components of the Dirac
operator. 

According to the principles of bulk quantization, we introduce three
fields to complete the quartet $\Phi$, $ \Psi$, $ \Aphi$, $ \B$.
A nilpotent BRST operator is defined by $s\Phi = \Psi$, $s\Psi =0$,
$s\Aphi = \B$ and $s\B = 0$.  $\Phi$ and $\Psi$ have the same canonical
dimension 1.  $\Aphi$ and $\B$ have dimension 2.  This can be deduced
from the fact that $\B$ will appear in combination with the equations
of motion for $\Phi$ in a field 
\begin{eqnarray}  \label{Hdef}
\H = \B + \bar D^2 \bar\Phi + V'(\Phi).  
\end{eqnarray}
With the additional knowledge that the ghost time $t$ has dimension
$-2$ and that chiral superfields must be integrated on $d^2 \theta$,
which adds one to the canonical dimension and general superfields must
be integrated on $d^4 \theta$, which adds two, we define the following
$s$-invariant and supersymmetric action of ghost number zero with only
positive dimension coupling constants:
\begin{eqnarray}\label{WZbrs}
&& \int dt\, d^4 x\, d^2 \theta\; s(\Aphi \partial_t \Phi)
	+ \int dt\, d^4 x\, d^2 \bar\theta\; s(\bar\Aphi \partial_t \bar\Phi)
\nonu \\
&&+ \int dt\, d^4 x\, d^4 \theta\; s( \bar\Aphi \H + \Aphi \bH ) \nonu \\
&&+ \int dt\, d^4 x\, d^2 \theta\; Ms(\Aphi \H)
+ \int dt\, d^4 x\, d^2 \bar\theta\; Ms(\bar\Aphi \bH)  .
\end{eqnarray}
$M$ is an additional mass parameter introduced by bulk quantization. 
Physical quantities will not depend on $M$, because the $M$-dependence
will appear through   a kernel.

We will shortly justify that  the field $\phi$ only appears through
$\partial_t \phi$ and $\H$ in eq.~(\ref{WZbrs}) by the 
invariance under bulk time reversal.
The ghost independent part of this action is:
\begin{eqnarray} \label{WZstoc}
&& \int dt\, d^4 x\, d^2 \theta\; \B \partial_t \Phi
        + \int dt\, d^4 x\, d^2 \bar\theta\; \bB \partial_t \bar\Phi
\nonu \\
&&+ \int dt\, d^4x\, d^4\theta\; \bigl(2 \B\bB + \B ( \ephi ) + \bB ( \bephi )
\bigr)
\nonu \\
&&+ \int dt\, d^4 x\, d^2 \theta\; M\bigl( \B^2 + \B(\ephi)\bigr)\nonu \\
&& + \int dt\, d^4 x\, d^2 \bar\theta\; M\bigl( \bB^2 + \bB (\bephi)\bigr)  .
\end{eqnarray}
Variation with respect to the superfields $\B$ and $\bB$ gives the following
matrix equation:
\begin{equation} \label{WZmat}
\pmatrix{ \partial_t \Phi \cr
             \partial_t \bar \Phi }=
        \pmatrix{ M    &\bar D^2\cr
                    D^2  &M }
 \pmatrix{  \bar D^2\bar\Phi+V'(\Phi) +2\B\cr
                         D^2\Phi+V'(\bar\Phi) +2\bar\B }
\end{equation}
These equations look like stochastic equations for the superfields $\Phi$
and $\bar\Phi$ with driving noises $\B$ and $\bB$, but this interpretation
(that would hold true for a genuine scalar field theory) does not stand out. 
In particular, $\B$ mixes with the physical field $\Phi$.
The action~(\ref{WZstoc}) is quadratic in the field $\B$, but as in the
fermionic case, the quadratic term $\int \B\bB$ contains space derivatives.
The propagator is a matrix with $\Phi-\Phi$ and $\B-\Phi$ propagations.

From the superfield $\Phi$, three fields can be defined
by taking the $\theta=\bar\theta=0$ value (denoted by the symbol $|$) of
the fermionic derivatives of $\Phi$:  
\begin{eqnarray}
A(x,t)&=&\Phi  (x,t,\theta,\bar\theta)     |,
\quad \varphi_\alpha(x,t)=D_\alpha \Phi  (x,t,\theta,\bar\theta) |,\nonu\\
\quad F(x,t)&=&D^2 \Phi  (x,t,\theta,\bar\theta)     |.
\end{eqnarray}
The equivalent fields for the $\B$ superfield are simply marked by a
subscripted $\B$.  
The detailed structure of eq.~(\ref{WZmat}) in components fields is:
\begin{eqnarray} \label{WZbos}
\pmatrix{ \partial_t A \cr
             \partial_t \bar F }&=&
        \pmatrix{ M    & 1 \cr
                    \Box  &M }
 \pmatrix{  \bar F +V'(A) +2 A_\B \cr
        \Box A +V''(\bar A) \bar F + g \bar\varphi\bar\varphi +2\bar F_\B }, \\
\pmatrix{ \partial_t \varphi_\alpha \cr
             \partial_t \bar \varphi^{\dot\alpha} }&=&
        \pmatrix{ M    & \partial _{\alpha\dot\alpha} \cr
                    \partial ^{\dot\alpha\alpha} &M }
 \pmatrix{  (\dirac\bar\varphi)_\alpha+V'(A)\varphi_\alpha +2 (\varphi_\B)_\alpha \cr
             (\dirac\varphi)^{\dot\alpha}+V'(\bar A)\bar\varphi^{\dot\alpha}
                         +2\bar(\varphi_\B)^{\dot\alpha} }. \label{WZfer}
\end{eqnarray}
Eq.~(\ref{WZbos}) must be completed by its complex conjugate to give the
evolution of $\bar A$ and $F$.  This set of equations is by
construction invariant under supersymmetry.  Notice that it turns the
field $F$, which is just an auxiliary field in the usual approach, into
a propagating field.  Therefore, if one keeps the auxiliary field to
maintain supersymmetry in a linear realization, the bulk formulation
implies its propagation.

In the heuristic language of stochastic quantization,
the $t$ evolution of $A$, $F$ and $\varphi$ is driven by the
four-dimensional equations of motion and $F_\B$, $A_\B$ and $\varphi_\B$,
their respective noises.  The form of the kernels that multiply the
combination of equations of motion and noises is dictated by the requirement
that the $t$ evolution is compatible with supersymmetry and the respective
dimensions of the fields and their equations of motion.  We believe that
eqs.~(\ref{WZbos},\ref{WZfer}) open a way toward a non-perturbative
definition of supersymmetric theories.

The ghost completion of eq.~(\ref{WZstoc}) is very easy to derive from
eq.~(\ref{WZbrs}).  It is such that the full action is BRS-invariant. 
For correlators at equal time $t$ of fields $\Phi$ and $\B$ only, which
are sufficient to determine the $S$-matrix of the model, the ghosts can be
integrated out without destroying supersymmetry.
It is useful to write the ghost equations of motion:
\begin{eqnarray} \label{WZfan}
\pmatrix{ \partial_t \Psi \cr
             \partial_t \bar  \Psi }&=&
        \pmatrix{ M    &\bar D^2\cr
                    D^2  &M }
 \pmatrix{  \bar D^2\bar\Psi+V''(\Phi)\Psi \cr
                         D^2\Psi+V''(\bar\Phi)\bar\Psi},\\
\label{WZafan}\pmatrix{- \partial_t \Aphi \cr
             -\partial_t \bar  \Aphi }&=&
        \pmatrix{ M    &\bar D^2\cr
                    D^2  &M }
 \pmatrix{  \bar D^2\bar\Aphi+V''(\Phi)\Aphi \cr
                         D^2\Aphi+V''(\bar\Phi)\bar\Aphi}.
\end{eqnarray}

We must now precise the symmetry with respect to the reversal of the bulk
time $t$, which implies that the $s$-invariant action is really of the
form~(\ref{WZbrs}). Moreover this symmetry  implies  the stability of
the theory under radiative corrections.
It reads as follows:
\begin{eqnarray}\label{tsym}
t&\to &-t,\quad \Phi(x,t)\to \Phi(x,-t), \nonu\\
\Psi(x,t)&\to& \Aphi(x,-t), \quad \Aphi(x,t) \to -\Psi (x,-t),\\
B(x,t) &\to& -H(x,-t).\nonu
\end{eqnarray}
This discrete symmetry explicitly depends on the dynamics, due to the
appearance of the classical equations of motion in $\H$.
Notice that 
the renormalization of vertices with only external lines of $\Phi$ and $\B$
fields does not involve the ghosts, since the latter have retarded
propagators and do not contribute to closed loops.
This symmetry completely justifies the expression~(\ref{WZbrs}) of the  action.
We also see that the equations of motion~(\ref{WZmat}) are invariant under
the symmetry~(\ref{tsym}).

\section{Supersymmetric Gauge Theories}
For theories with a gauge symmetry, quantization with a ghost time involves
an additional subtlety.  The topological symmetry in five dimension must be
disentangled from the gauge symmetry.  This problem has been recently
solved by introducing a second BRST symmetry $w$ which anticommutes with
$s$.  The theory is then defined as being invariant under both $s$ and $w$.
Due to the topological character of the $s$-symmetry, the cohomology of $s$
is empty.  The observables are defined from the cohomology of $w$.
The full complement of fields necessary to realize both symmetries is
described in~\cite{BaGrZw00} for the Yang--Mills theory.

This general construction should work also in the supersymmetric case.
Technical difficulties however arise and the subject deserves a separate
publication~\cite{BaBe01}.  In the following, we restrict ourselves to
present the basic features in the simpler SQED case and sketch the basic
problems.

The supersymmetric Maxwell field is described by a real superfield
$V$ of dimension 0, with a chiral gauge parameter $\Lambda$ acting
by $V\to V + i(\Lambda - \bar\Lambda)$.  The gauge invariant field
strength is the chiral superfield $W_\alpha = \bar D^2 D_\alpha V$ and its
complex conjugate $\bar W_\da$.
The equation of motion is $D^\alpha W_\alpha = 0$, which is equivalent
to $D^{\dot\alpha} \bar W_{\dot\alpha} =0 $ by the Bianchi identity
and is of dimension 2.  But we cannot simply combine the study of gauge
theories of~\cite{BaGrZw00} with a superspace formalism.  In superspace,
the gauge parameter becomes a full multiplet, which gives algebraic gauge
variations to some components of the superfield $V$.  It is therefore
necessary to verify that all the components of the superfield $V$,
either physical fields or pure gauges, have well-defined propagators.

The topological BRST symmetry is similar to the one for the Wess-Zumino
model, but with additional terms that account for the gauge symmetry in
superspace.  A chiral superfield $X$ must be introduced. It is the
supersymmetric generalization of the gauge field component $A_5$ that
plays a key role for the bulk quantization of the genuine Yang--Mills
theory~\cite{BaGrZw00,BaZw00}.  We introduce
chiral superfields $C$ and $\phi$ as additional ghost companions to $X$,
$\fX$:
\begin{eqnarray}
sV &=& \fV + i(C - \bar C), \quad s\fV = -i ( \phi - \bar \phi), \nonu \\
sC &=& \phi, \quad s\phi = 0, \nonu \\
sX &=& \fX + \partial_t C,\quad  s\fX = -\partial_t \phi. 
\end{eqnarray}
We have the antighosts $\aV$ and $\aphi$ and the corresponding Lagrange
multipliers $\bV$ and $\eta$:
\begin{equation}\label{YMag}
        s \aV = \bV, \quad s \bV =0, \quad s \aphi = \eta, \quad s \eta = 0.
\end{equation}
We suppose that the full $w$ and $s$ BRS-like symmetries will require
additional fields, as they had to be introduced in the Yang--Mills
case~\cite{BaGrZw00,BaZw00}, but we do not go in these details now.

The Lagrangian will
be written with the combination of the equations of motion and noise
$\bV + D^\alpha \bar D^2 D_\alpha V $.
We  expect to obtain the following $t$-evolution of the superfield $V$:
\begin{equation}\label{SQEDn}
        \partial_t V = \bV + D^\alpha \bar D^2 D_\alpha V+i(X-\bar X).
\end{equation}
The last term is necessary to determines the $t$ evolution over the
supersymmetric generalization of gauge orbits and justifies the
introduction of the superfield $X$.

Now comes the difficulty. The later equation must be derived from  a
Lagrangian that is  built from the $s$-variation of the integral
in superspace of $\aV$ times the desired variation of $V$ given in
eq.~(\ref{SQEDn}),
\begin{equation} \label{QEDLa1}
 \int dt\, d^4x\, d^4\theta\; s\,\Bigl( \aV( \partial_t V
- D^\alpha \bar D^2 D_\alpha V +\bV -i(X-\bar X) ) \Bigr) .
\end{equation}
Variation with respect to an unconstrained $\bV$ yields
equation~(\ref{SQEDn}), but for them to have the correct physical
interpretation, we need the linear constraint on $\bV$, $D^2 \bV=0$ and
$\bar D^2 \bV=0$.  Without such a constraint, the auxiliary $D$-field
would eventually appear as a propagating scalar, since its equation of
motion from eq.~(\ref{SQEDn}) is of the type $\partial_t D = \Box D +
B_{D} $, where $ B_{D} $ is the scalar auxiliary component in $B_V$.  
In the constrained linear multiplet, this higher component of the field
is equal to the Laplacian acting on the scalar field of the multiplet.
Similarly, the equation of motion of the photino would be 
$\partial_t \lambda = \Box \lambda + B_\lambda$.  The constraint on
$\bV$ insures that $B_\lambda $ is equal to the Dirac operator acting on
the fermionic component of the supermultiplet, which yields the correct
interpretation.  However, variation with respect to a constrained field
would not produce eq.~(\ref{SQEDn}), but some superspace derivative of
it.  In this case, the equation would act only on the field strength
$W_\alpha$, letting the gauge variant part of the field $V$ without a
definite evolution.  

These difficulties seem to be related to the fact that the detailed balance of
degrees of freedom between fields of different types is not so simple
to obtain as in the Wess--Zumino model.  The effective number of
degrees of freedom of the gauge field $V$ is reduced by the gauge
symmetry, but this mechanism is not available for the antighost and
noise.  In fact, these are quite similar to the canonical momenta in
Hamiltonian gauge field theory, which are constrained by Gau\ss'
law.  The equivalent superfield constraint is satisfied by the field
equations and defines a linear superfield. (Let us recall that a a
linear superfield $l$ is a real superfield satisfying $D^2 l = 0$ and
$\bar D^2 l=0$; its independent components are a scalar, a Majorana
spinor and a divergenceless vector field.)  As in the quantization of
the Hamiltonian gauge field theory, one would wish to express this
constraint as the equation obtained from the variation of the time
component of the connection.  In superspace, the connection is not an
ordinary vector and what play the role of the time-like component of
the connection is a chiral superfield, which transforms as the time
derivative of the chiral superfield gauge parameter. 

\section{Conclusion}

In this letter, we have shown that the quantization with an additional
time can be defined while maintaining supersymmetry on the example of the
Wess--Zumino model. Since this
method of quantization is expected to be more profound from a non
perturbative point of view, this is a progress in view of
further tests for supersymmetry. In a separate publication, we
will explain how our result can be extended to the case of
supersymmetric gauge theories.

{\bf Acknowledgements}: We thank Luis Alvarez-Gaum\'e for very useful
discussions concerning this work.  The research of Laurent
Baulieu was supported in part by DOE grant DE-FG02-96ER40959.

\end{document}